\documentclass[10pt, conference, compsocconf]{IEEEtran}
\usepackage[utf8]{inputenc}

\usepackage{graphicx}
\graphicspath{ {./images/} }
\usepackage{float}
\usepackage{amsmath}
\usepackage{listings}
\usepackage{courier}

\usepackage{url}

\usepackage[table,x11names,dvipsnames,table]{xcolor}

\begin{document}
\title{Self-Sovereign Identity for IoT environments:\\ A Perspective}

\author{
\IEEEauthorblockN{
Geovane Fedrecheski\IEEEauthorrefmark{1},
Jan M. Rabaey\IEEEauthorrefmark{4},
Laisa C. P. Costa\IEEEauthorrefmark{1},\\
Pablo C. Calcina Ccori\IEEEauthorrefmark{1},
William T. Pereira\IEEEauthorrefmark{1},
Marcelo K. Zuffo\IEEEauthorrefmark{1}}
\IEEEauthorblockA{
\IEEEauthorrefmark{1}Interdisciplinary Center on Interactive Technologies, Polytechnic School,\\
University of Sao Paulo, Brazil\\
\IEEEauthorrefmark{4}Berkeley Wireless Research Center, Electrical Engineering and Computer Science Department,\\
University California, Berkeley, US}\\
\{geovane, laisa, mkzuffo\}@lsi.usp.br,
pcalcina@ime.usp.br,
williamtpereira@gmail.com,
jan\_rabaey@berkeley.edu
\thanks{This research was partially funded by CAPES.}
}

\maketitle

\begin{abstract}
This paper analyses the concept of Self-Sovereign Identity (SSI), an emerging approach for establishing digital identity, in the context of the Internet of Things (IoT).
We contrast existing approaches for identity on the Internet, such as cloud-based accounts and digital certificates, with SSI standards such as Decentralized Identifiers (DIDs) and Verifiable Credentials (VCs).
To the best of our knowledge, this is the first thorough comparison of these approaches.
The benefits and challenges of using DIDs and VCs to identify and authenticate IoT devices and their respective users are discussed. In the end, we establish that SSI, with its owner-centric, privacy-aware and decentrailized approach, provides a viable and attractive option for secure identification of IoT devices and users.

\end{abstract}
\IEEEpeerreviewmaketitle

\section{Introduction} 

The Internet was developed as a research project to interconnect computers \cite{leiner2009brief}. Protocols like TCP/IP, developed as open standards, allowed computers to connect in a global scale.
However, even after the world-changing impacts the Internet had on society over the last decades, it has no pervasive, privacy-preserving, and easy to use mechanism to manage digital identities.

Where human activity is involved, a common abstraction is to use accounts, i.e. digital records, often containing personally identifiable information (PII), that are protected by a password and saved on a webserver. 
Although this method has been working for several decades, it has many security drawbacks, such as the use of weak passwords \cite{taneski2019systematic} and the potential for privacy violation.
Furthermore, the manual approach of password-protected accounts makes it unsuitable to machine-to-machine interactions, a common scenario in the IoT.

More automated solutions can be achieved by using Public Key Certificates (PKCs) that bind names to public keys \cite{kohnfelder1978towards}. Widespread use of PKC, however, is limited to organizations, due to the complexity of current methods. For instance, while websites usually prove their identities to web browsers using certificates, users do not use certificates in the same way, i.e. to prove their identity to the website. 
Moreover, existing standards were not designed for privacy, as evidenced by the use of real names in known certificate formats such as PGP \cite{RFC4880openPGP} and X.509 \cite{RFC5280x509}.
To aggravate the situation, the assignment of unique names often require centralized architectures, which is inadequate for distributed IoT applications.


A recent development towards online identification of users, organizations, and devices has been referred to as ``Self-Sovereign Identity'' (SSI).
The basic premise of SSI is that subjects should own and control their own identity, instead of having it stored and managed by a third party.
This approach brings several benefits, including enhanced privacy, control, and decentralization.
Two new standards are being proposed to realize SSI, namely, Decentralized Identifiers (DIDs) and Verifiable Credentials (VCs) \cite{W3C2019decentralized, W3C2019verifiable}. While DIDs focus on cryptographic identification, VCs provide a means for privacy-aware and authenticated attribute disclosure.

In this paper we analyze existing approaches to identity in the Internet, such as X.509, PGP \cite{RFC4880openPGP}, and SSI. We present a detailed comparison focusing on the data models used to represent identity across different standards. Finally, we discuss what are the benefits of using SSI in the Internet of Things, and identify challenges that must be overcome. 












\section{Self-Sovereign Identity}

Self-Sovereign Identity is an approach in which subjects are in full control of their own digital identities \cite{allen2016path}. SSI is analogous to offline identifiers, which are carried by the owner (within a physical wallet), but contrasts with current digital identity solutions, which are either based on accounts or digital certificates, and have privacy and centralization issues.

While initially proposed by members \cite{allen2016path} of online communities, a formal definition of SSI was released recently \cite{ferdous2019search}. Considering an identity to be composed of an identifier associated with a set of name-value attributes, the full self-sovereign identity of an individual is the collection of all identities (i.e. identifiers and attributes) that span a range of decentralized domains, such that the individual is in full control of these identities \cite{ferdous2019search}.
As digital privacy concerns have been growing in recent years, interest in SSI has intensified. 
This led to the definition of a set of technical specifications to implement SSI, which we describe below.

\subsection{Decentralized Identifiers}
Digital identifiers so far have been either centralized or non-resolvable. For example, Uniform Resource Locators (URLs), which can be used to resolve HTML documents, usually depend on domains names assigned by ICANN\footnote{Internet Corporation for Assigned Names and Numbers - https://www.icann.org/}, a centralized authority. On the other hand, unique, user-generated identifiers such as UUIDs cannot be used to resolve associated metadata.

To address this, a new specification for Decentralized Identifiers (DIDs) is being developed with the support of the W3C \cite{W3C2019decentralized}. The DID has the following syntax: \texttt{did:btcr:abcdefgh12345678}. The \textit{did} prefix is mandatory, and colons are used to separate a \textit{method definition} and a \textit{method-specific id}. A \textit{method} is a specific set of rules for working with DIDs (the example above uses the Bitcoin method), and the format of the \textit{id} depends on that method. An open directory of different DID methods is available for public access and open for new submissions\footnote{https://w3c-ccg.github.io/did-method-registry/}.

Each DID is associated with a DID Document (DDo) that contains the DID itself along with public keys, service endpoints, and other metadata. The public key is used to authenticate and encrypt messages, while the endpoint provides a way to message the entity that controls that DID. To control a specific DID, a subject just have to own a private key associated with public keys in the DDo.

A common storage mechanism for DDos are Blockchains, from which they can be resolved using the referred DID. 
On the other hand, in some cases individuals may not want to publish their DIDs, e.g. to avoid identity correlation. In this case, the special \textit{peer} DID method can be used.
Thus, DIDs are unique identifiers that can be resolved to DID Documents, and they allow the establishment of an end-to-end secure channel. What DIDs do not provide, however, is a means for entities to prove claims (attributes) about themselves.

\subsection{Verifiable Credentials}
\begin{figure}[ht]
\centering
\includegraphics[width=.37\textwidth]{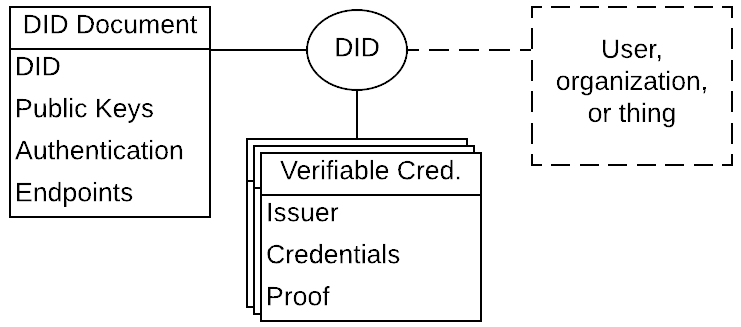}
\caption{A DID is the link between a DDo and a set of VCs, much like a primary key can link different tables in a database.
This allows a subject associated with a DID to prove its identity.
}
\label{fig:did}
\end{figure}

\begin{figure}[ht]
\centering
\includegraphics[width=.48\textwidth]{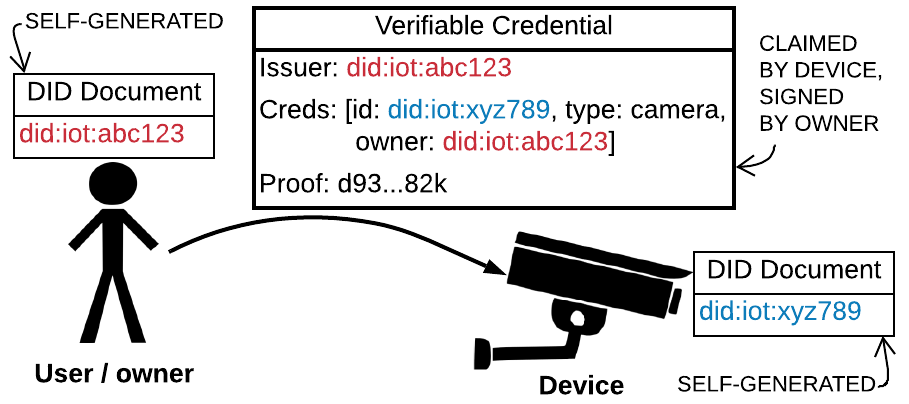}
\caption{An owner-centric scenario using SSI. Each subject generates its own DDo, while the VC is issued by the device owner.}
\label{fig:example}
\end{figure}

Verifiable Credentials (VCs) is a W3C recommendation for portable and provable claims about a subject. For instance, a person may claim to have the name Alice, and a device may claim to be of type Camera. 
The relationship among DIDs and VCs is shown in Figure \ref{fig:did}.
All VCs refer to the DID of the subject to which they have been assigned (e.g. an IoT device). VCs also contain the DID of its issuer along with a cryptographic proof.
This allows a subject to present a VC to a verifier, which can then resolve the DDo of the issuer (and therefore its public key) from a public ledger, e.g. a Blockchain, and check the authenticity of the VC.
Figure \ref{fig:example} shows a use case where a user issues a VC to a device.

A major incentive for SSI is privacy, therefore VCs are expected to be private and stored in a personal wallet, to be shared only when necessary.
To further improve privacy, the VC specification supports zero-knowledge proofs, i.e. 
a cryptographic technique ``where an entity can prove to another entity that they know a certain value without disclosing the actual value'' \cite{W3C2019verifiable}. 


\subsection{Decentralization, privacy, and layered authentication}


Public key cryptography can be used to derive a shared secret over an insecure channel \cite{diffie1976new}. However, a known problem is how to trust the origin of the public key.
To solve this, a signed certificate that binds a name to a public key was proposed \cite{kohnfelder1978towards}. Two common standards for digital certificates are X.509 \cite{RFC5280x509, RFC5755attribute} and Pretty Good Privacy (PGP) \cite{RFC4880openPGP}.
Although they differ in details, both follow the original definition in which names
are tied to public keys and signed by a third party \cite{kohnfelder1978towards}.

A crucial challenge faced by certificate-based solutions was ensuring the uniqueness of the names. The most common solution to this was to rely on centralized architectures. For example, the name on the \texttt{subject} field in X.509 must be enforced by a global authority, and the PGP id uses the name of a person plus her email address, which ultimately depends on DNS, which is centralized as well.

More recently, the emergence of Blockchain technology allows decentralized consensus for choosing unique names.
One problem, however, is that solutions based on certificates put sensitive information in the identifier, which compromises the privacy of certificate holders, and therefore might not be suitable for storage in public, immutable ledgers.

An approach to solve this is to limit the exposure of PII on the ledger by only writing anonymous information to it, e.g., public keys. In particular, this approach enables public key storage and lookup, which can be used to create a confidential and non-repudiable channel.
Higher-level abstractions can then be used to implement authentication, since the attributes necessary to authenticate users are usually application-specific.

This is the solution that results from combining the DID and VC specifications. 
Containing only pseudonymous information, such as public keys and service endpoints, DID Documents can be used to establish a cryptographically secure channel between two entities. After the confidential channel is created, the entities can exchange VCs, according to the levels of trust necessary to each application. 
In other words, while DIDs are lower-level and pseudonymous, VCs are application-specific and can be used to authenticate attributes such as name or device type. 
Finally, it is worth noting that as each DID is usually a high-entropy random string, name collisions actually stop being a concern.

\section{Data models for digital identity} 
\label{sec:tools}

This section provides analysis and comparisons of existing data models for digital identity.
We start by discussing the limitations of password-based accounts, and then proceed to compare data models based on public key cryptography.

\subsection{Accounts}
The most basic method to identify subjects in computer systems is the \textit{account}: a digital record, usually composed of at least a user name and a password, that identifies a user. Accounts are commonly stored in a server controlled by the service provider.
For example, popular IoT vendors require that a device owner have a cloud-based account, so that she can use this virtual identity to configure her devices.

While accounts have been used for decades in a variety of systems, they are among the most primitive solutions for digital identities. Among the problems related to account-based authentication are privacy and the use of passwords.
With respect to privacy, issues arise because the user is forced to store plaintext PII in a third-party system. Regarding passwords, the literature indicates common problems such as password reuse and difficulty to enforce strong passwords, and points that the most widespread solution is the use of ``recommendations'' \cite{taneski2019systematic}, which depends on human factors and are difficult to enforce.

\begin{table*}[]
\centering
\caption{Comparison of standardized data models for digital identity.}
\label{tab:comp}
\rowcolors{2}{gray!25}{white}
\begin{tabular}{l|l|l|l|ll}
\rowcolor[HTML]{C0C0C0} 
\multicolumn{1}{c|}{\cellcolor[HTML]{C0C0C0}}                               & \multicolumn{1}{c|}{\cellcolor[HTML]{C0C0C0}\textbf{PGP}}                                                                                  & \multicolumn{2}{c|}{\cellcolor[HTML]{C0C0C0}\textbf{X.509}}                                                                                                                                                                                                              & \multicolumn{2}{c}{\cellcolor[HTML]{C0C0C0}\textbf{Self-Sovereign Identity}}                                                                                                                                                                                         \\ \cline{2-6} 
\rowcolor[HTML]{C0C0C0} 
\multicolumn{1}{c|}{\cellcolor[HTML]{C0C0C0}}                               & \multicolumn{1}{c|}{\cellcolor[HTML]{C0C0C0}\textbf{PGP Key}}                                                                              & \multicolumn{1}{c|}{\cellcolor[HTML]{C0C0C0}\textbf{\begin{tabular}[c]{@{}c@{}}Public Key Certificate \\ (PKC)\end{tabular}}}              & \multicolumn{1}{c|}{\cellcolor[HTML]{C0C0C0}\textbf{\begin{tabular}[c]{@{}c@{}}Attribute Certificate \\ (AC)\end{tabular}}} & \multicolumn{1}{c|}{\cellcolor[HTML]{C0C0C0}\textbf{\begin{tabular}[c]{@{}c@{}}DID Document \\ (DDo)\end{tabular}}}                     & \multicolumn{1}{c}{\cellcolor[HTML]{C0C0C0}\textbf{\begin{tabular}[c]{@{}c@{}}Verifiable Credential \\ (VC)\end{tabular}}} \\
\textbf{Goal}                                                               & \begin{tabular}[c]{@{}l@{}}Prove control of public \\ keys and identifier \\ (plus optional attributes)\\ Publish public keys\end{tabular} & \begin{tabular}[c]{@{}l@{}}Prove control of public \\ keys and identifier \\ (plus optional attributes)\\ Publish public keys\end{tabular} & \begin{tabular}[c]{@{}l@{}}Prove possession of \\ attributes\end{tabular}                                                   & \multicolumn{1}{l|}{\begin{tabular}[c]{@{}l@{}}Prove control of identifier\\ Publish public keys and \\ service endpoints\end{tabular}} & \begin{tabular}[c]{@{}l@{}}Prove possession of \\ attributes\end{tabular}                                                  \\
\textbf{Identifier}                                                         & Name and Email                                                                                                                             & Qualified Name                                                                                                                             & Same as PKC                                                                                                                 & \multicolumn{1}{l|}{Method-specific DID}                                                                                                & Same as DDo                                                                                                                \\
\textbf{\begin{tabular}[c]{@{}l@{}}Uniqueness\\ of identifier\end{tabular}} & \begin{tabular}[c]{@{}l@{}}Global \\ authority (DNS)\end{tabular}                                                                          & Global authority (CA)                                                                                                                      & Same as PKC                                                                                                                 & \multicolumn{1}{l|}{\begin{tabular}[c]{@{}l@{}}Ledger consensus /\\ Random number gen.\end{tabular}}                                    & Same as DDo                                                                                                                \\
\textbf{Public Key(s)}                                                      & 1 primary, N subkeys                                                                                                                       & 1                                                                                                                                          & n/a (points to PKC)                                                                                                         & \multicolumn{1}{l|}{N}                                                                                                                  & n/a (points to DDo)                                                                                                        \\
\textbf{Attribute(s)}                                                       & Attributes field                                                                                                                           & Extensions field                                                                                                                           & Attributes field                                                                                                            & \multicolumn{1}{l|}{-}                                                                                                                  & subjectCredential field                                                                                                    \\
\textbf{Endorsement}                                                        & Signature of many peers                                                                                                                    & Signature of a CA                                                                                                                          & Signature of a CA                                                                                                           & \multicolumn{1}{l|}{\begin{tabular}[c]{@{}l@{}}Self-signed (optional)\\ Indirect through VC\end{tabular}}                               & Signature of an Issuer                                                                                                     \\
\textbf{\begin{tabular}[c]{@{}l@{}}Service \\ endpoints\end{tabular}}       & -                                                                                                                                          & -                                                                                                                                          & n/a                                                                                                                         & \multicolumn{1}{l|}{Yes}                                                                                                                & n/a                                                                                                                        \\
\textbf{\begin{tabular}[c]{@{}l@{}}Semantic \\ schemas\end{tabular}}        & -                                                                                                                                          & -                                                                                                                                          & -                                                                                                                           & \multicolumn{1}{l|}{Yes}                                                                                                                & Yes                                                                                                                       
\end{tabular}
\end{table*}

\subsection{Models based on public key cryptography}

Pretty Good Privacy (PGP) \cite{RFC4880openPGP} was created to allow individuals to prove a binding between a public key and an identifier, the latter being composed by a real name and an email address. This binding, along with optional attributes and signatures, is stored in a document called a PGP Key. Conceived as a distributed solution, individuals in the PGP scheme can sign the keys of other individuals, so as to give an endorsement that they are who they say they are, i.e. they are not impersonating someone or using a fake id. This scheme of peer signatures is often referred to as the Web of Trust.

X.509 Certificates, created by the X.500 working group, defines a format for Public Key Certificates (PKC) that binds public keys to qualified names \cite{RFC5280x509}. PKCs are widely used in the Internet to authenticate domain names and protect communications. Although technically nothing prevents peer-to-peer signature of X.509 certificates, the vast majority of its usage is under centralized architectures, in which a trusted authority signs the certificate to make it trustworthy. 
Finally, in certain cases it is useful to have a separate document that, instead of having public key, contains only a name associated with signed attributes. To meet this demand, X.509 proposed a new standard called Attribute Certificate (AC), which contains no public key, but links to a PKC through its \textit{subject} field \cite{RFC5755attribute}.

Finally, as previously mentioned, Self-Sovereign Identity is a novel approach that uses Decentralized Identifiers \cite{W3C2019decentralized} and Verifiable Credentials \cite{W3C2019verifiable} to prove possession of identifiers and attributes, respectively.

\subsection{High-level comparison}

The following paragraphs compares models used by the PGP, X.509, and SSI standards, according to Table \ref{tab:comp}.

\paragraph*{Goal} Both PGP Keys and PKCs are used to publish and prove control of public keys that are tied to identifiers. Also, in these approaches, attributes can be provided either in the same document as the public keys (PGP Key and PKC), or, in the case of X.509, in a separate document (AC). On the other hand, documents in the SSI paradigm have decoupled goals: DDos are be used to prove control of an identifier and to provide a means for establishing a secure communication; and VCs are used to prove possession of attributes.


\paragraph*{Identifier (and uniqueness)} While PGP and X.509 use names and other identifiers that depend on centralized entities, in SSI the identifiers are completely decentralized and can be auto-generated, for example by using strong random number generators. Not only this enables easy global uniqueness, but the pseudonymous characteristic of DIDs also enhances privacy, when compared to previous approaches based on real names or email addresses. Pseudonymous identifiers are also more suited for IoT, since devices do not have names or email addresses by default.

\paragraph*{Public Key(s)} PKCs are limited to only one public key, while PGP Keys and DDos can have many. 
PGP still differs from DDos as the former uses a primary key that is tied to an identifier and allows more subkeys to be included, while the latter support multiple public keys without assumptions other than the key type, which usually encodes its purpose, e.g. sign or encrypt.

\paragraph*{Attribute(s)} Both PGP Keys and X.509 certificates support arbitrary attributes, either via PKC extensions or dedicated ACs. In self-sovereign identity, a DDo does not support attributes in order to stay anonymous. Instead, all PII is handled only by VCs, which are private by default.

\paragraph*{Endorsement(s)} PGP Keys can be signed by one or more peers, but X.509 certificates and VCs can only be signed by a single issuer.
DDos are not signed by external entities, and may be self-signed. When a DDo is written to a ledger, however, the transaction will be signed, which can be used to attest the validity of the DDo. Another way of proving endorsement over a DID is to check the signature of a VC associated with that DID. If the VC is signed by a trusted issuer, the DID can be trusted.
Furthermore, with respect to who can sign the endorsements, technically it can be anyone, but there are philosophical differences. X.509, for example, was devised to work within a centralized architecture, where only trusted authorities can sign certificates. On the other end of the spectrum, PGP expects peer-to-peer signatures, which ultimately creates a Web of Trust.
Finally, VCs does not make strong assumptions on the network structure, although decentralized approaches, especially the ones based on Blockchain, may be favorable.

\paragraph*{Service endpoints} a novelty introduced by DDos is the association of a built-in mechanism to reach the owner of a public key. This facilitates the establishment of secure interactions between peers, from web to IoT environments.

\paragraph*{Semantic schemas} only SSI-based data models allow extensibility through semantic annotations over JSON documents. 
The main reason for this is that these technologies only became popular after X.509 and PGP were developed.

\subsection{Public key distribution}

\begin{table}[]
\centering
\caption{Comparison of data models for key distribution.}
\label{tab:keydist}
\rowcolors{2}{gray!25}{white}
\begin{tabular}{l|l|l|l}
\rowcolor[HTML]{C0C0C0} 
\multicolumn{1}{c|}{\cellcolor[HTML]{C0C0C0}\textbf{}}                                & \multicolumn{1}{c|}{\cellcolor[HTML]{C0C0C0}\textbf{Raw Pub Key}} & \multicolumn{1}{c|}{\cellcolor[HTML]{C0C0C0}\textbf{PKC}} & \multicolumn{1}{c}{\cellcolor[HTML]{C0C0C0}\textbf{DDo}}         \\
\begin{tabular}[c]{@{}l@{}}Associates key material \\ to metadata\end{tabular}        &                                                                   & X                                                         & X                                                                \\
Privacy: no PII disclosed                                                             & X                                                                 &                                                           & X                                                                \\
\begin{tabular}[c]{@{}l@{}}Key rotation does not \\ requires re-signing\end{tabular}  & n/a                                                               &                                                           & X                                                                \\
Serialization formats                                                                 & \begin{tabular}[c]{@{}l@{}}Binary\\ Base64\end{tabular}           & \begin{tabular}[c]{@{}l@{}}DER\\ PEM\end{tabular}         & \begin{tabular}[c]{@{}l@{}}JSON-LD\\ JWT\end{tabular} \\
Semantic schemas                                                                      &                                                                   &                                                           & X                                                                \\
\begin{tabular}[c]{@{}l@{}}Decentralized: user \\ generates the artifact\end{tabular} & X                                                                 &                                                           & X                                                                \\
\begin{tabular}[c]{@{}l@{}}Decentralized: user \\ carries the artifact\end{tabular}   & X                                                                 & X                                                         & X                                                                \\
Service endpoint                                                                      &                                                                   &                                                           & X                                                               
\end{tabular}
\end{table}

\begin{table}[]
\centering
\caption{Comparison of data models for attributes.}
\label{tab:creds}
\rowcolors{2}{gray!25}{white}
\begin{tabular}{l|l|l|l}
\rowcolor[HTML]{C0C0C0} 
\textbf{}                                                                                & \textbf{PKC}                                      & \textbf{AC}                                       & \textbf{VC}                                           \\
\begin{tabular}[c]{@{}l@{}}Signed attributes \\ about a subject\end{tabular}             & X                                                 & X                                                 & X                                                     \\
\begin{tabular}[c]{@{}l@{}}Key rotation does not \\ requires re-signing\end{tabular}     &                                                   & X                                                 & X                                                     \\
\begin{tabular}[c]{@{}l@{}}Identifier differs from \\ key material\end{tabular}          & X                                                 & X                                                 & X                                                     \\
\begin{tabular}[c]{@{}l@{}}Attributes decoupled \\ from key material\end{tabular}        &                                                   & X                                                 & X                                                     \\
Selective disclosure                                                                     &                                                   &                                                   & X                                                     \\
Zero-knowledge proofs                                                                    &                                                   &                                                   & X                                                     \\
Delegation                                                                               &                                                   & X                                                 &                                                       \\
Revocation                                                                               & X                                                 & X                                                 & X                                                     \\
Serialization formats                                                                    & \begin{tabular}[c]{@{}l@{}}DER\\ PEM\end{tabular} & \begin{tabular}[c]{@{}l@{}}DER\\ PEM\end{tabular} & \begin{tabular}[c]{@{}l@{}}JSON-LD\\ JWT\end{tabular} \\
Semantic schemas                                                                         &                                                   &                                                   & X                                                     \\
\begin{tabular}[c]{@{}l@{}}Decentralized: user \\ carries the artifact\end{tabular}      & X                                                 & X                                                 & X                                                     \\
\begin{tabular}[c]{@{}l@{}}Decentralized: Verifier \\ decoupled from Issuer\end{tabular} &                                                   &                                                   & X                                                    
\end{tabular}
\end{table}

An important aspect in the design of systems based on asymmetric encryption is the data model used to support key distribution. In the following, we compare three approaches, as shown in Table \ref{tab:keydist}: Raw Public Key, Public Key Certificates, and DID Document.

\paragraph*{Raw public key} this is the simplest approach, and consists in having a public key shared as a raw array of bytes, often encoded in some ascii-compatible format, such as base64. Although this approach is decentralized and discloses no personal information, it does not allow associated metadata.

\paragraph*{Public Key Certificate} as previously discussed, PKCs bind a name and other attributes to a public key, which allows subjects prove their identity. Created before privacy was a major concern, X.509 PKCs always carry PII in the main identifier, and may carry PII in other attributes. Finally, other drawbacks of PKCs include the imposition of specialized serialization formats (DER and PEM), tight coupling of keys and data (which makes key rotation more difficult), and a centralized architecture, i.e. the artifact is not self-generated.

\paragraph*{DID Document} DDos associate public keys to pseudonymous metadata, while also allowing key rotation without re-signing any associated metadata. 
The latter is possible because all signed metadata actually only lives in associated VCs.
An important difference to highlight is that DDos are not signed by third parties, thus they cannot authenticate the origin of a public key. If this is necessary, DDos can be composed with VCs to increase security.
DDos supports JSON-based serialization formats, which are available in most programming languages and platforms, and can benefit from publicly available semantic schemas.
As each user auto-generates their own DIDs and DDos, the management of the identifier is decentralized.
Finally, service endpoints in DDos provide a novel way for peers to securely establish secure channels.

\subsection{Attribute distribution}

Four out of the five previously described formats can be used to prove control over attributes: PGP Keys, Public Key Certificates, Attribute Certificates, and Verifiable Credentials. Since PGP Keys are less widely used, we only compare the latter, as shown in Table \ref{tab:creds}.

\paragraph*{Public Key Certificates} the encoding of attributes in PKCs leverages the X.509 PKC \textit{extension field}. Although the reuse of an existing format may seen advantageous in terms of compatibility, the whole certificate must be re-signed when a key is rotated, or when selective disclosure of attributes is necessary. 
An important drawback not mentioned so far is that it is impossible to disclose only a subset of the attributes in a PKC, without contacting the issuer for a new signature.

\paragraph*{Attribute Certificates} differing from PKCs, ACs contain a name and a list of attributes, but no public key, which simplifies key rotation. Finally, while ACs support delegation, in general they have the same drawbacks as PKCs.

\paragraph*{Verifiable Credentials} similar to an AC, a VC does not contain public keys, as it focus on binding identifiers to attributes. 
Among the novelties in the VC standard is the support for selective disclosure without contacting the issuer, which is realized using zero-knowledge cryptography. 
VCs also leverage JSON, a serialization format that is both human readable and lightweight to parse. VCs and can be further specialized into two formats: JSON Linked Data (JSON-LD)\footnote{https://json-ld.org/}, a format to serialize linked data; and JSON Web Token (JWT), a widely used format to express security claims\footnote{https://jwt.io/}.

\section{Benefits and Challenges of SSI for IoT}
\label{sec:benefits}

As the IoT continues to evolve, new paradigms that allow spontaneous machine-to-machine interactions started to appear \cite{rabaey2011swarm, costa2015swarm}. 
Necessarily decentralized, the future IoT will require users to be the root of trust of their devices, leading to an owner-centric IoT. As privacy concerns raise in importance, solutions that minimize personal data sharing become paramount. Full realization of these and other features will require novel, open, and secure standards for identity in the IoT.
The next paragraphs discuss aspects of self-sovereign identity that are likely to improve decentralized IoT security, while also pointing the factors that will require innovation to bring SSI to IoT, such as support constrained devices.


\subsection{Benefits}
The benefits of SSI for IoT, such as privacy and decentralization, are discussed below.

\paragraph*{Owner-Centric}
The user can be the root of trust of her devices. Once a user is the owner and controller of her identity, it is straightforward to create a network of devices that belong to her, for example by provisioning an ``owner=Alice'' credential to each device. One interesting consequence of this is that no third party is needed to enforce security and administration of devices, as the user herself will be able to do it. Note that in this approach devices can have their own identity as well, and may only use the owner attribute to facilitate the creation of trust relationships, i.e. devices that share the same owner can automatically trust each other.  

\paragraph*{Privacy-preserving}
Personal information is protected. By having the identity of owners and things stored locally, sensitive data that would otherwise be stored in a service provider will now live closer to the owner (usually in a digital wallet). While the user can choose to backup his data for various reasons, she will be able to do so in an encrypted way, as only she will possess the decryption keys. Users and devices will also get to choose with whom they share their credentials, and even be able to do so employing selective disclosure and zero-knowledge proofs techniques, further improving privacy.

\paragraph*{Decentralized}
No single-point of failure. While identity providers may have been a convenient way to authenticate users and devices so far, it is not clear what happens when a provider stops providing, e.g. when it goes out of business. In the self-sovereign approach, the user decides when her identity starts or stops being valid, and she will have similar controls over her devices. Finally, data breaches, information sharing without user consent, and other issues are minimized when identities are not stored in a high-value data silo that 
acts as a honeypot for
hackers.

\paragraph*{End-to-end security}
Communications between two endpoints are secure. By exchanging DID Documents and applying asymmetric cryptography, IoT devices can mutually authenticate, derive short-lived symmetric keys, send encrypted messages, and enforce non-repudiation. This approach can also be implemented in a transport-agnostic way, enabling secure communication even among different protocols.

\paragraph*{Layered authentication}
Separates cryptographic and application-specific authentication. In the former, two devices prove to each other that they are in possession of specific public keys, while in the latter the devices prove different attributes about themselves.
This approach allows endpoints to always be cryptographically protected, and leaves higher-level trust requirements to be handled at the application layer.


\paragraph*{Standardized and open approach}
Fosters interoperability and robustness.
Since both DIDs and VCs are being developed as open W3C specifications, companies and researchers are free to build solutions that are interoperable and rely on well-tested data models.


\paragraph*{JSON-based encoding} Using JSON enables more applications to handle data extracted from DID Documents and credentials, even if not originally designed to work with SSI.


\subsection{Challenges}
We now discuss some challenges to apply SSI in IoT environments.

\paragraph*{Constrained devices}
Fully adopting SSI means that devices need to be able to run asymmetric cryptography and cope with communication overhead of transmitting metadata, such as DID Documents and Verifiable Credentials.

\paragraph*{Asymmetric Cryptography}
SSI demands execution of encryption algorithms based on asymmetric keys, which can be challenging in devices with limited processing and energy resources. While authors points that constrained processors such as the  32-bits Cortex M0 are well equipped to execute Elliptic Curve Cryptography (ECC) \cite{kortesniemi2019improving}, the number of operations still must be controlled to avoid battery draining. A common tactic is to use long lived session keys that are less frequently updated, e.g. once a day.

\paragraph*{Communication overhead}
Depending on the communication protocol, the size of DDos and VCs may impose a barrier. For example, low energy protocols such as LoRA and BLE have maximum packet sizes of 222 and 244 bytes, respectively, while DDos and VCs easily achieve 500 bytes or more. Therefore, strategies such as compression, fragmentation, and infrequent document transmission, will be necessary. In extreme cases, SSI may not be possible at all, which will require proxy approaches \cite{lagutin2019enabling}.

\paragraph*{DID Resolution}
Higlhy constrained devices may not be able to connect to the Internet to download DID Documents at all. A possible solution is to create a local cache of known DIDs, either managed by the device itself or by its gateway. On the other hand, if both devices use peer DIDs, they can simply exchange their DIDs directly, which shifts the problem to securely delivering the DIDs in the first place.

\paragraph*{Software availability}
The SSI ecosystem is new and there is limited software available for embedded devices. Given the foundational importance of secure cryptographic algorithms and protocols, applications based on SSI should rely on existing libraries that encapsulate complexity and are well tested, which reduces the chances for vulnerabilities. 
Although reference implementations exist \cite{foundation2019aries}, they are focused on cloud and mobile use cases.
To fully incorporate SSI into IoT, portable and lightweight libraries tailored for constrained devices must be created and made widely available.


\section{Conclusion and perspective}
As the primary motivation for the development of the Internet was to remotely connect computers, the problem of secure identification of users and devices was left aside.
While identity solutions such as accounts and certificates were eventually developed, they feature critical issues such as weak passwords, lack of privacy, and centralization.
As it is common for systems to mature over time, as good (and bad) practices are learned, we argue that the Self-Sovereign Identity approach represents an important step forward in the area of digital identity.
Particularly in the context of the IoT, this paper showed how SSI can (1) empower owners to have full control over both their identities and their devices, (2) improve privacy by decoupling pseudonymous and sensitive identity records, and (3) allow decentralized identity management by reducing the dependency on third parties.
As for the next steps, the realization of SSI in the IoT will demand implementations that are optimized for constrained devices, both for cryptographic operations and low-power communication. Furthermore, wide adoption of SSI will depend on the availability of open software libraries to manipulate DIDs and VCs in IoT devices.
To conclude we argue that, if adopted, SSI may significantly benefit security and privacy of IoT applications, and potentially enable new use cases, such as those that involve cross-owner decentralized interactions.



\bibliographystyle{IEEEtran}
\bibliography{IEEEabrv,references}
\end{document}